\documentclass[
]{ceurart}

\usepackage{listings}
\begin{document}

\copyrightyear{2025}
\copyrightclause{Copyright for this paper by its authors. Use permitted under Creative Commons License Attribution 4.0 International (CC BY 4.0).}
\conference{AEQUITAS 2025: Workshop on Fairness and Bias in AI | co-located with ECAI 2025, Bologna, Italy}

\title{Modeling Fairness in Recruitment AI via Information Flow}


\author[1]{Mattias Brännström}[%
orcid=0000-0003-3113-2631,
email=mattias.brannstrom@umu.se,
]
\cormark[1]
\address[1]{Umeå University, Universitetstorget 4, 901 87 Umeå, Sweden}

\author[1]{Themis Dimitra Xanthopoulou}[%
orcid=0000-0003-2914-0472,
email=themis.xanthopoulou@umu.se,
]

\author[1]{Lili Jiang}[%
orcid=0000-0002-7788-3986,
email=lili.jiang@umu.se,
]

\cortext[1]{Corresponding author.}


\begin{abstract}
Avoiding bias and understanding the real-world consequences of AI-supported decision-making are critical to address fairness and assign accountability. Existing approaches often focus either on technical aspects, such as datasets and models, or on high-level socio-ethical considerations—rarely capturing how these elements interact in practice. In this paper, we apply an information flow-based modeling framework to a real-world recruitment process that integrates automated candidate matching with human decision-making. Through semi-structured stakeholder interviews and iterative modeling, we construct a multi-level representation of the recruitment pipeline, capturing how information is transformed, filtered, and interpreted across both algorithmic and human components. We identify where biases may emerge, how they can propagate through the system, and what downstream impacts they may have on candidates. This case study illustrates how information flow modeling can support structured analysis of fairness risks, providing transparency across complex socio-technical systems.
\end{abstract}


\maketitle

\section{Introduction}
Artificial Intelligence (AI) is rapidly finding its way into all areas of society, shaping decision-making processes across personal, business, public, and private sectors\cite{theodorou2020towards,rubenstein2021acquiring}. Its integration into existing organizations brings significant challenges, particularly in addressing fairness, bias, and accountability in socio-technical systems. These issues are especially pressing as AI systems increasingly take part in high-stakes decisions, affecting diverse stakeholders with varying degrees of vulnerability. While awareness of the potential dangers of automated decision-making is increasing and being recognized in regulation, prominently in the EU AI Act, there is a lack of tools and methodologies to fully address these challenges. 

Current efforts to address these challenges can be broadly categorized into two approaches with their own perspectives strengths and weaknesses when it comes to characterizing and addressing these challenges. 
The first, technical perspective, focuses on aspects like bias detection in datasets, fairness metrics, and model explainability \cite{mehrabi2021survey,ntoutsi2020bias}. These methods prioritize measurable model-level statistics, such as fairness definitions derived from confusion matrices, but often fail to account for the structural and organizational dynamics that shape how AI systems operate in practice \cite{selbst2019fairness,weinberg2022rethinking}. This disconnect risks what Selbst et al. have called the \textit{framing trap} — where system-level fairness or compliance is wrongly inferred from component-level properties \cite{selbst2019fairness}. This risks reducing fairness and other metrics to disconnected mathematical constructs without a clear connection to real-world implications of algorithmic decisions.

The second, socio-technical perspective is found in high-level guidelines, Fundamental Rights Impact Assessments (FRIA), legal and socio-technical frameworks~\cite{VealeBorgesius2021,EC2024Framework}. These approaches emphasizes high-level ethical principles, societal impacts, and stakeholder accountability \cite{rain,cobbe2021reviewable}. However, while tackling the socio-technical side of the gap, they frequently lack tools for analyzing or appreciating the internal mechanics of technical systems and their embedding in real-world workflows. In lack of representation of the technical side these approaches risk treating the technical system as a singular opaque entity and introduce methodological black boxes. Without fine-grained tools for connecting technical and social elements in a manner that directly connects to the socio-technical impacts the gap remain and these methods become limited in explaining how biases emerge and propagate into stakeholder impacts~\cite{green_algorithmic_2020,selbst2019fairness}.

The abstraction gap forming in between the socio-technical and technical spheres prevent a holistic understanding of how technical details and stakeholder effects interact. The problem is not merely one of methodological focus, but one of lacking bridge structures across the gap. While efforts exist on both sides of the gap to capture something of the other side, significant challenges remain~\cite{metcalf2021algorithmic, green_algorithmic_2020, weinberg2022rethinking, selbst2019fairness}. 


To address this gap, we have developed the Information Flow Model (IFM) methodology, which models socio-technical decision systems as flows of information. IFM represents information states as sites and the transformations between them as channels, mapping how inputs are reshaped into outputs and decisions. This provides a form of structural transparency: it makes visible how technical operations and human judgments interact within the broader organizational process, and how biases and responsibilities can propagate through these interactions to produce downstream impacts. In this way, IFM models bridge the gap between technical and socio-technical approaches, as both model-level metrics and higher-level organizational and stakeholder considerations can be connected and traced across the same structural substrate.

\textbf{Contribution.} This paper presents a real-world case study applying the IFM methodology to a recruitment process that combines AI-supported and human decision-making. We make three contributions:
\begin{enumerate}
    \item We demonstrate how IFM enables joint modeling of technical components (such as automated matching systems) and surrounding social processes (such as recruiter interpretation and client selection).

    \item We show how structuring the decision process as flows of information through sites and transformations supports systematic identification of where biases may be introduced, how they propagate, and what impacts they may have on stakeholders.

    \item We illustrate how the IFM bias impact analysis clarifies socio-technical structures of decision systems, makes information dependencies explicit, and traces how roots of discrimination arise and propagate across both algorithmic and human steps when the method is performed in a participatory setting.
\end{enumerate}

Taken together, the case highlights IFM’s potential as a pragmatic tool for system-level analysis and design in real-world contexts.


\textbf{Structure of the Paper} The rest of the paper is structured as follows: Section \ref{sec:framework} gives a brief technical description of IFM. Section \ref{sec:methodology} gives a short methodological overview. Section \ref{sec:results} presents the results of the use-case study, presenting information flow models and analysis of bias and impact. 
We discuss the outcomes of the case-study and framework as well as directions for further study and development in Section \ref{sec:discussion}.





\section{Background}

The challenges of AI fairness and accountability are not defined by a single gap but by multiple separations forming this abstraction gap. These include divergences of method, abstraction, and perspective, which fragment the field into disconnected approaches that struggle to speak to one another~\cite{green_algorithmic_2020,weinberg2022rethinking}.

Bridging the gap thus requires a combination of holistic, structural and situated perspectives. A structural perspective connects decision-making, dependencies and consequences. A holistic perspective gives an interconnected whole, not be just a collection of details, and ensures that the we are not creating small framing traps ~\cite{selbst2019fairness,VealeBorgesius2021}. Finally, challenges and consequences are not high level principles but situated in the context of particular stakeholders~\cite{weinberg2022rethinking}.  

\subsection{Related work and IFM}

Impact assessments such as AIAs and FRIAs foreground governance, rights, and accountability, but stop at the surface of system documentation and do not address how technical operations actually unfold \cite{VealeBorgesius2021,EC2024Framework}. Bias taxonomies and fairness metrics dissect sources of error and discriminatory correlations, offering diagnostic precision yet failing to capture how distortions propagate through socio-technical workflows \cite{mehrabi2021survey,ntoutsi2020bias,makhlouf2021applicability}. Causal models do try to trace the dependencies within information, data and outcomes, but they mostly work within the technical frame such as individual datasets rather than the full socio-technical frame~\cite{loftus2018causal,su2022review,makhlouf2020causalsurvey}. Process modeling traditions, including BPMN, UML and FRAM, contribute structural descriptions of workflows and, in FRAM’s case, variability analysis across socio-technical functions~\cite{salehi2021modeling,chinosi2012bpmn,recker2006bpmn}. While these approaches do provide a structural account of workflows and decision processes, they typically lack of holistic closure. They might model structures piecemeal without semantics creating a continual connected whole, focusing on communicating some functional details but not the full structure of decision making~\cite{chinosi2012bpmn,recker2006bpmn}. These frameworks also remain disconnected from fairness concerns and lack mechanisms for tracing how information dependencies and biases link technical operations to stakeholder impacts.

Our proposed IFM methodology aims to address these separations not by replacing existing approaches, but by providing a structural substrate in which they can connect. At its core, IFM models decision-making as networks of information sites and channels, where each transformation detail it's information dependencies and outputs. The IFM graph form a mesh which might be arbitrarily detailed down to technical components but still span over the same socio-technical frame. Through paths can be traced downstream to stakeholder impacts or upstream to structural causes. This structural mesh ensures a combination of holistic closures and capture of technical properties.

While we will here present IFM in particular, the real intended strength of IFM  is less as a stand-alone tool and more as a connective framework, providing the scaffolding through which technical, organizational, and stakeholder perspectives can meet in the middle, bridging the abstraction gap.

\subsection{Situatedness and Participatory Approaches}

Situatedness is essential in order to understand real consequences of unfairness such as particular stakeholder impacts and the challenges involved to tackle these risks~\cite{selbst2019fairness,weinberg2022rethinking}. This means we will not gain much from modeling 'a recruitment system' or 'a classifier' as all such will be different, and embedded differently as soon as you look at a real actual socio-technical system. Essential to approach the real situated system is the inclusion of stakeholders and their contextual perspective \cite{participatoryabeba,participatoryturn}.

The IFM method employed in this paper and in general uses participatory methods, interviews and workshops as the primary method to obtain a situated model capturing the dynamics and processes as described and understood by those who interact with it, more than using system specification or other objective descriptions. 

Participatory approaches can be understood in several ways: as acknowledgment of the democratic right to participate \cite{participatoryabeba}; as recognition that no single stakeholder can adequately represent all perspectives or knowledge systems \cite{participatoryabeba}; and as appreciation of the impact that inclusive processes can have, ranging from product improvements to improved community outcomes \cite{participatoryturn}. 

The landscape of participation is characterized by different levels and types \cite{participatoryturn}, each serving distinct purposes and offering varying degrees of stakeholder involvement. However, participation also presents notable challenges that must be carefully navigated. Critically, \emph{participation washing} - where the participation of stakeholders are used merely to justify an otherwise obtained design or model - does not constitute effective participation \cite{participatoryabeba}.

Understanding how to achieve effective participation is essential for meaningful engagement. It is argued that for participatory design, risk mitigation, or audit processes, participants who do not necessarily possess specialized technical skills must first develop sufficient understanding of the sociotechnical system \cite{participatoryabeba}. Researchers advocate for providing participants with contextual material rather than technically dense documentation \cite{situatedparticipation}. Additionally, beyond incorporating localized knowledge, some scholars suggest that participants should develop an understanding of how AI systems are embedded within broader sociotechnical contexts and how these systems interact with human users \cite{participatoryturn}.

With IFM we partly flip this perspective. The goal of the IFM modeling process is to understand how potential AI components are embedded within the broader socio-technical system and it is not necessarily so that anyone possess this understanding at the start of the modeling process. Instead each stakeholder knows their own role, be it a technical role, a user or other stakeholder. The main aim of the IFM participatory process is to let each stakeholder communicate their situated perspective and thereby shape their part of the decision environment. An initial lack of understanding of the whole is not just expected but informative, showing the role of potential transparency or explainability mismatches. As modeling progresses the developed model shift towards being the object of the study, a role it can take just to the degree is properly capture the situated perspectives of the involved stakeholders. 
\section{The Framework} \label{sec:framework}

Studying how decision-making occurs in socio-technical systems requires a clear and systematic representation of how information flows through the decision process. The Information Flow Methodology (IFM) offers a structured way to describe these processes, enabling the identification of potential biases and the evaluations of how outcomes affect stakeholders.
The IFM model was developed loosely inspired by the works of Barwise \cite{barwise1995logic, barwise1997information} and uses similar components.
    
    This section provides a conceptual overview of the IFM and its core components (Section \ref{sec:IFMdescription}), followed by descriptions on the two structured properties - bias and impact, in IFM (Section \ref{sec:biasInIFM}). A more in depth treatment of the IFM method can be found in the IFM Methodological guide~\cite{ifm2025guide}.

    \subsection{An Overview of the IFM Model}\label{sec:IFMdescription}

    The IFM Model provides a detailed description of the decision-making process by specifying the information available for each decision and the resulting outcome. This specification enables a thorough analysis of the factors that might influence a decision outcome and how the resulting outcomes might, in turn, impact other decisions.

    The IFM consists of three primary components:
    \begin{enumerate}
        \item \textbf{Sites:} Represent the states or repositories of information within a decision process, such as inputs, intermediate results, or outputs.
        \item \textbf{Channels:} Represent the transformations, operations, or decisions that move information from one site to another.
        \item \textbf{Networks:} Represent sites and channels connected to form directed graphs, capturing entire decision processes from initial inputs to outcomes.
    \end{enumerate}
    
    Any decision process that begins with some initial information and ends with a decision can be described using these elements. The decision process itself constitutes a network, each sub-decision a \textbf{channel}, and the inputs, output decision, and any intermediate result would be \textbf{sites}.
    

    Following this conceptual overview, we extend the model with a formal definition to enable more rigorous analysis.

    \subsubsection{Formal Definition of the IFM Model} \label{sec:ifm_formal}

    A Network $N = <S, C, T_S, T_C, \mathcal{T}, \mathcal{R}>$ where $S$ is a set of Sites representing sources of information, $C$ is a set of Channels representing directed transformations between sites. $T_S: S \rightarrow T$ and $T_C: C \rightarrow T$ are functions assigning \emph{types} to sites and channels respectively. $\mathcal{T} = (T, \leq_T)$ is a type system where $T$ is a set of type classes and $\leq_T$ is a partial ordering defining subtype relationships. $\mathcal{R}$ is a set of inference rules determining properties and relationships based on types.

    A Site $S_i \in S$ represents a source of information on which a decision can be made. While sites are unique in $N$, the information they contain can be further described by $T_S$ assigning them one or more types. Types will be further described below and are not unique to a site.
    A Channel $C_j \in C$ is a directed transformation between one or more input sites $In(C_j) \subseteq S$ with one or more output sites $Out(C_j) \subseteq S$. Like sites, channels are unique in $N$ but can be assigned channel types via $T_C$ or inferred types via $R$.

    A valid network $N$ forms a directed acyclic graph (DAG) where:
    \begin{itemize}
    \item A site in $S$ can be the output of at most one channel in $C$.
    \item The output of a channel cannot serve as an input to an upstream channel (no cycles).
    \item $In(N) = \{S_i \in S \mid S_i \text{ is not an output of any } C_j \in C\}$.
    \item $Out(N) = \{S_o \in S \mid S_o \text{ is not an input to any } C_j \in C\}$.
    \item $Mid(N) = S \setminus (In(N) \cup Out(N))$ \quad (intermediate outputs of $N$).
    \end{itemize}
    
    A network $N$ can be seen as an abstract channel $N: In(N) \rightarrow O$ where $O \subseteq Out(N)$. Similarly, any channel $C$ can be further defined as a network. This allows a network $N$ to be nested within another larger network $N'$ where it participates as a channel. This nesting provides the possibility of abstraction, as parts of large or complicated networks can be abstracted as channels representing sub-networks. It also allows causal reasoning about the properties that networks have as channels.
    
    \subsection{Alternatives} \label{sec:alternatives}

    Alternatives are used to represent uncertainty or options within the formal description. Alternatives are defined as a set of possible values, possibly including the absence of a value (denoted $\empty$). Alternatives will be denoted either by specifying the set of alternatives prefixed by a question mark "?", for instance $?\{A1, A2\}$, which denotes that $A1$ or $A2$ might apply, or by applying the $?$ operator to a symbol $A? = ?\{A, \empty\}$, denoting that $A$ might not be present. While alternatives could be applied to any property such as sites, channels or types, we will primarily handle them on a network level. A network or sub-network containing alternatives represents multiple possible configurations, where each configuration corresponds to one specific network without alternatives.


    \subsection{Bias and Impact}\label{sec:biasInIFM}
    
    In the IFM model, both \emph{bias} and \emph{impact} are treated as structured properties of the modeled decision process. They reflect different but connected forms of misalignment that may affect stakeholders or violate regulatory norms and obligations.

    \textbf{Defintion of Bias}, in this context, refers to a systematic deviation from the described function of the decision process. Such biases might result in decreased accuracy through false positives and false negatives and the effects might be limited or disproportionate to specific subsets of input, such as favoring particular groups.
    Bias may propagate \emph{downstream} through dependent channels and sites. If its influence is halted or unused, the propagation path ends.
    Once bias is introduced in the information flow it might spread \emph{downstream}, towards outputs. Any downstream channel that depends on a biased site can be treated as a potential proxy for that bias. 
    Conversely, if biased information is not used further or the bias is controlled, the propagation path terminates. 
    This makes it possible to trace the scope and effect of bias and mitigating efforts across the network, in a transparent and formally bounded way.
    
    \textbf{Definition of Impact} here refers to the real-world normative consequences or risk thereof of decisions, especially unintended harms affecting stakeholders or sub-groups. 
    Impacts strongly relate to bias as bias are sources of unintended decisions and information in the information flow, but harmful impacts can be caused also by the system working as intended if the design have not taken risk-mitigation into account. As impacts occur at the downstream ends of the DAG, understanding their causes requires tracing their paths \emph{upstream} towards contributing decisions and information.
    Where the flow of bias downstream and the tracing of impact upstream intersect, the bias may a contributing cause to the impact. Since some impacts only occur for particular subgroups, when analyzing contributing paths we might need to follow identifiers or proxy variables connected to this sub-group through the information flow. This determines if the proxy identifier contributes to the impact.

\section{Method} \label{sec:methodology}

To analyze fairness-relevant risks in an AI-supported recruitment process, we applied the Information Flow Model (IFM) methodology\cite{ifm2025guide}. 
The approach enables structured reasoning about how information is represented, restructured, and filtered as it moves through human and algorithmic components of decision-making. Our goal in this case study was to construct such a model grounded in semi-structured interviews, and use it to identify potential sources of bias and trace their downstream impacts.

The recruitment process under analysis includes both manual and automated steps: recruiters interpret client needs, generate or refine job descriptions, use AI-based matching tools to identify candidates, and conduct interviews before forwarding recommendations to the client. These activities involve multiple actors and complex collection and use of information --- making it a particularly good case for IFM modeling. 

The methodology was applied in five main steps, adapted to the use-case:

\begin{enumerate}
    \item \textbf{Sketching the initial model.}  
    Three semi-structured interviews were conducted with stakeholders (two recruiters, one developer). During the interviews, the two interviewers independently produced informal sketches of the information flow based on the participants' descriptions of how decisions were made and what information was used. This interview process is intentionally dynamic and focus on the areas which seems most relevant to both the participants and modelers. Interviews also capture more general views of the stakeholders perspective which help greatly to ground further analysis.
    
    \item \textbf{Formalization and refinement.}  
    The sketches were consolidated and refined channel per channel. Channels were reviewed to ensure valid inputs, outputs, and other formal requirements. This stepwise refinement leads to the creation of a formal IFM model. Studying each channel obtained from the previous step one at a time helps to some degree in reducing modeler bias on presumed understanding of the whole flow. If the step by step refinement runs into problems this might prompt the need for additional interviews or reflect actual inconsistencies in need of exploration.
    
    \item \textbf{Semantic typing and actor assignment.}  
    Each site and channel was annotated with semantic types (e.g., list, sublist, unstructured description, etc.) and assigned responsible actors (client, recruiter, AI system). Chanels are similarly typed into filtering, ranking, and abstraction and assigned to sub-networks with semantic labels. Depending on what information is present it might be useful to describe it further, such as what features are actually present in a dataset. In other types a more abstract definition can be used to capture unstructured information. This flexibility is used to facilitate a holistic perspective which might be difficult on a fixed abstraction level. 
    
    \item \textbf{Identification of potential bias.}  
    After understanding types and channel semantics, the model was again reviewed, channel for channel, for transformations that could introduce problems or bias. This step also involved taking existing mitigation mechanisms into account and noting them down. Ambiguities and inconsistencies from the previous modeling process can be marked here as potential sources of bias. Key on this step is that there is no detection of actual bias possible here, from a structural account alone, instead it is the envelope of \emph{potential} bias which is mapped out - which is structurally dependent.
    
    \item \textbf{Analysis and tracing of Impact paths.}  
    Finally, in Section \ref{sec:discussion} we analyze whether and how identified biases can propagate downstream to affect candidate outcomes. This is done by establishing if there is a path between a potential source of discrimination or bias through the information flow all the way to a stakeholder outcome. If such a path can be described, then there is structural risk for this impact to materialize.
\end{enumerate}

For a more in depth description on IFM, it's parts and the methodology, see the IFM Methodological Guide \cite{ifm2025guide}. 
The results of the first four of these steps are presented in detail together with figures in the following section, followed by step five in the following analysis and discussion section.

\section{Results}\label{sec:results}
This section presents the results from the interview study and following structural analysis.

As per Steps 1 and 2 of the IF methodology three semi-structured interviews were held with use-case stakeholders. Two with recruiters and one with an internal software developer. 

\paragraph{A summary of the recruitment process: } From these interviews, a general picture of a workflow emerged. With the help of outsourced recruiters, the client with vacancies to fill describes the vacancy to recruiters who create a job description and technical requirements. 
The recruitment company then in a \emph{Sourcing} phase employs different ways to collect lists of potentially interested and suitable candidates based on requirements. Recruiters search platforms such as LinkedIn for candidates and advertise the job description on various websites. The recruitment company owns a database where candidates register themselves if they are interested in being recruited, or get registered if they apply for a position from another website. To search the database of candidates, the recruitment firm has an AI Matching tool that matches the characteristics of candidates with the job requirements to provide a ranked list of candidates. This list improves the efficiency of the recruitment assignment since there can be too many candidates in the database and searching through filters does not provide holistic comparisons. 

After the Sourcing phase begins the Screening phase. The Recruiter reviews the AI list to ensure the results are valid. Depending on the number of open positions for the role, the recruiter selects top-ranked candidates from the AI matching results and additional candidates from other recruitment sources to proceed to the next step. The selected candidates are contacted by the recruiter, and those who pass this initial screening advance to a more thorough interview to assess their skills. The final list of candidates are presented to the client who selects the final candidates to get job offers.

\subsection{Construction of the Information Flow}

\begin{figure*}[ht]
    \centering 
    \includegraphics[width=1.0\linewidth]{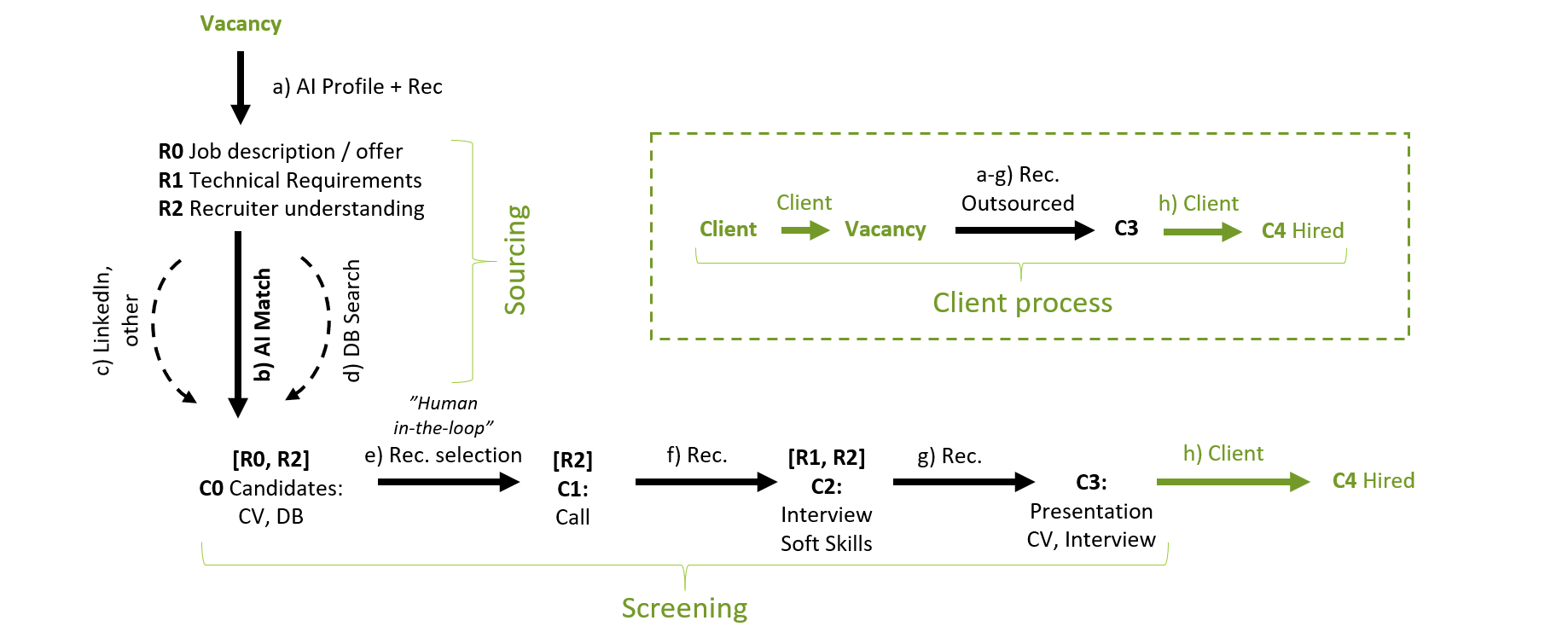}
    \caption{Overview of the recruitment process information flow corresponding to Table \ref{tab:transitions}. The included dashed green box displays the client process in where the channels a-g) is embedded. The channel AI Match is further detailed in Figure \ref{fig:ai_match}.}
    \label{fig:rec}
\end{figure*}

The information flow model was created following the methodology which is summarized in Section \ref{sec:methodology}. \textbf{Step 1} of the methodology is about charting out the information flow which underlies the decision process. This sketching was performed independently by two interviewers during the interview process. These sketches --- from each interviewer and from each interview --- are then compiled together info a collection of channels. \textbf{Step 2} is about refining these channels so that they conform to the formal definitions of the IFM, thereby creating the formal model. The models created are presented as two figures and two tables, together describing three nested levels of decision making. Figure \ref{fig:rec} describes the decision process of the client, and embedded in this the decision flow of the outsourced recruitment process. Table \ref{tab:transitions} gives more detail on each channel of the recruitment process marked in the figure as $(a)-(h)$.

\begin{figure*}[ht]
    \centering 
    \includegraphics[width=1.0\linewidth]{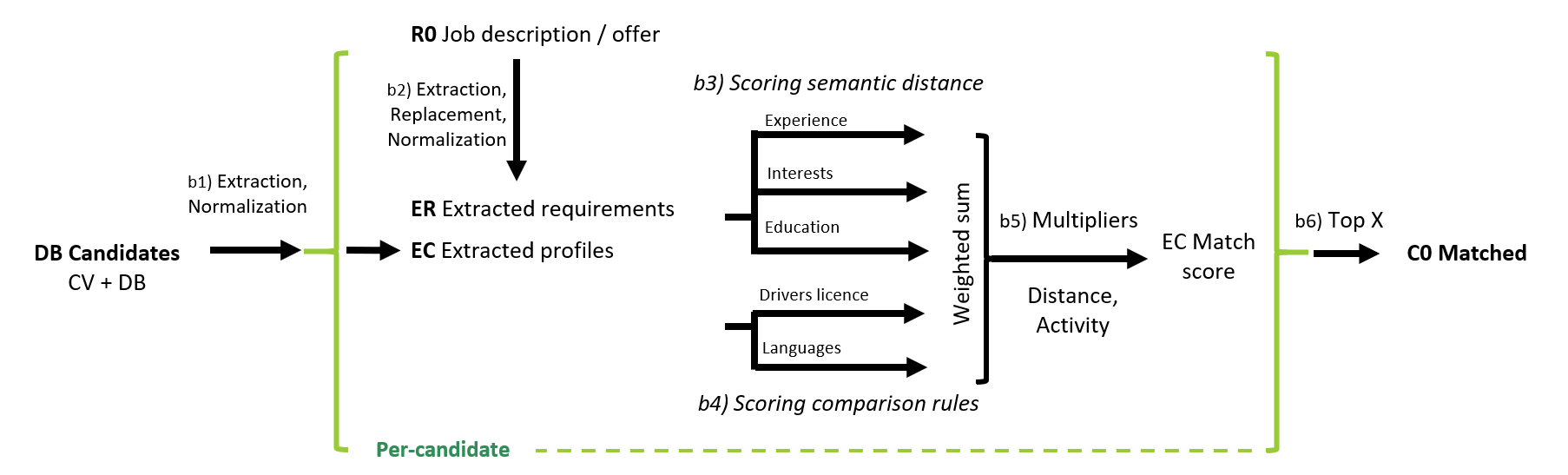}
    \caption{More detailed overview of AI Matching, corresponding to Table \ref{tab:ai_match_breakdown}.}
    \label{fig:ai_match}
\end{figure*}

Several AI tools are used at various points of the recruitment process, most prominently creating of job-descriptions in channel $(a)$ and automatized candidate matching in channel $(b)$. In the interviews we explored the AI Matching system, channel $(b)$ in more depth and we therefore present it in detail in Figure \ref{fig:ai_match} and Table \ref{tab:ai_match_breakdown} covering channels $(b1) - (b6)$.
The bare formal model is contextualized in \textbf{Step 3} by assigning semantic types, roles and actors to sites, channels and sub-networks. The most relevant of this information is shown in the tables and and figures.

As highlighted by the multi-abstraction-level representations in Figures \ref{fig:rec} and \ref{fig:ai_match} the information flow model can be made very detailed or very abstract. The relevant level of analysis depends on the questions which are to be explored. In this case these three levels adequately describe the most relevant for our analysis of the recruitment process and their embedded context.

\begin{table*}[t]
\centering
\resizebox{0.95\textwidth}{!}{%
\begin{tabular}{clllcll}
\toprule
\textbf{\#} & \textbf{Transition} & \textbf{Actor} & \textbf{Sub-Network} & & \textbf{Potential Bias} & \textbf{Impact} \\
\midrule

a & $\text{Client} \xrightarrow{\text{AI Profile + Rec}} R0,R1,R2$ 
& Recruiter & Client process & & Interpretation, Normalization & \textbf{I1, I2} (client, rec) \\

b & $R0, \text{Candidate DB} \xrightarrow{\text{AI Match}} C0_b$ 
& AI & Sourcing & & Opacity, see Table \ref{tab:ai_match_breakdown} & \textbf{I3} (location) \\

c & $R0, R1, X \xrightarrow{\text{LinkedIn, other}} C0_c$ 
& Recruiter & Sourcing & & Interpretation &  \\

d & $R2, \text{Candidate DB} \xrightarrow{\text{DB search}} C0_d$ 
& Recruiter & Sourcing & & Interpretation ($R2$) &  \\

e & $[R0, R2], C0_b, C0_c, C0_d \xrightarrow{\text{Rec. filter}} C1$ 
& Recruiter & Screening & & Presentation &  \\

f & $C1 \xrightarrow{\text{Call}} [R2], \text{Impression} \xrightarrow{\text{Rec. filter}} C2$ 
& Recruiter & Screening & & Interpretation ($R2$) &  \\

g & $C2 \xrightarrow{\text{Interview}} [R1, R2], \text{Soft skills} \xrightarrow{\text{Rec. filter}} C3$ 
& Recruiter & Screening & & Interpretation ($R2$) &  \\

h & $C3 \xrightarrow{\text{Client Selection}} C4$ 
& Client & Client process & & Interpretation, Presentation & \textbf{I1} (client) \\

\bottomrule
\end{tabular}
}
\caption{Recruitment system transitions with actor roles, biases, and downstream impacts}
\label{tab:transitions}
\end{table*}

\begin{table*}[t]
\centering
\resizebox{0.95\textwidth}{!}{%
\begin{tabular}{clllll}
\toprule
\textbf{\#} & \textbf{Transition} & \textbf{Inputs} & \textbf{Operation} & \textbf{Bias, (Mitigation)} & \textbf{Impact} \\
\midrule

b1 & $\text{DB} \xrightarrow{\text{Extract}} EC$ & DB Candidate CVs & Feature extraction & Abstraction, (Normalization) & \\

b2 & $\text{?R0, Vacancy} \xrightarrow{\text{Extract}} ER$ & Job Description (R0) & Feature extraction & Abstraction, Opacity, (Normalization) & \\

b3 & $EC, ER \xrightarrow{\text{Semantic distance}} S1$ & Extracted profiles, requirements & Semantic similarity scoring & Misalignment & \\

b4 & $EC, ER \xrightarrow{\text{Rule comparison}} S2$ & Structured features & Rule-based matching & Rigidity & \\

b5 & $S1, S2 \xrightarrow{\text{Weighted sum}} S3$ & Scored components & Aggregation via weighted sum & Subjective heuristics & \\

b6 & $S3 \xrightarrow{\text{Multiplier}} S4$ & Aggregated score + context & Heuristic adjustment & Opacity, Subjective heuristics & \textbf{I3}  (location)\\

b7 & $S4 \xrightarrow{\text{Top X}} C0_b$ & Final scores & Thresholding / ranking & Score opacity, Exclusion & \\

\bottomrule
\end{tabular}
}
\caption{Breakdown of AI Match process (zoom-in on transition \textbf{b}) with potential biases and impacts}
\label{tab:ai_match_breakdown}
\end{table*}

\subsection{Analysis of Biases and Mitigation}

By considering each channel one by one, we, as \textbf{Step 4} of the methodology, consider what possible biases could take part of the type of transformation the channel represents. Some channels, particularly within the AI Matching system are designed with \emph{mitigation} of bias in mind, and these are also noted. We collect the information on potential biases in Table \ref{tab:transitions} and Table \ref{tab:ai_match_breakdown}. In addition, we note some of these biases below.

\subsubsection{Abstraction and Interpretation Biases}

In each channel where unstructured or semi-structured information is to be abstracted and represented there is a potential for information loss and distortion through abstraction. This type of abstraction occurs at several stages of the process starting with a) when the Client's vacancy is to be re-expressed as a job description with appropriate technical requirements. While abstraction by itself can cause systematic biases by missing features or systematic disregard of potentially important features, this becomes of particular concern when interpretation is involved. Since this decision process is to a large degree one of Recruiter judgment such interpretation can not be, and should not be, entirely avoided. However, with interpretation comes potential for bias. Biased interpretation can concern on one side information of candidates, and on another --- more systemic --- level information from the client. We are highlighting the interpretation involved by tracking the comparison to $R2$, the recruiters personal understanding of the client requirements as this is the main contrast in each of the more informal filtering steps.

\subsubsection{Presentation Bias}

At several steps of the process, it is likely that the way information, otherwise similar, is presented of included in the process affects how it is received and treated. These are marked as occasions of \emph{presentation} bias in Table \ref{tab:transitions} and \ref{tab:ai_match_breakdown}. Of particular mention is that candidates sourced from different pathways (AI Matching, DB-search, LinkedIn, personal contact, etc.) might be given different weight. This is in particular an issue since the sourcing paths themselves are mainly dependent on different information and might thereby shift the recruitment process towards or against more biased grounds of selection. 
Another type of presentation bias exist as the handoff between the AI Match system and the user --- the Recruiter --- where the user is presented a ranked list of results. The presentation of the ranked list hides the distance in score between the highest and lowest scored possibly creating a false impression of difference.

\subsubsection{Transparency and Opacity}

Recruiters lack access to the internal criteria and weighting used by the AI system, which may lead to overconfidence or misinterpretation of rankings. As mentioned above, the absence of numerical scores hides potential similarity between candidates. It also hides the working mechanism of the matching from the user and prevents them to form an accurate mental model of it's behavior. In interviews, recruiters expressed that they primarily saw the tool as a search tool --- while the developers primarily saw it as a ranking tool. This difference ultimately reflects a different view on how it operates. Beyond merely the results, the recruiters were also unaware that the tool discards their inputed job description $R0$ if there is a job description available from the client, here captured as \emph{Vacancy}. Effectively this means aspects of the recruiters own understanding $R2$ which influences $R0$ is prevented from influencing the outputs of the AI Matching. We can see this as a \emph{opacity bias} as it is a property causing an alternative information flow (excluding $R0$ and $R2$) contrary to user expectations --- thereby affecting the output. This is marked as $?R0$ in Table \ref{tab:ai_match_breakdown} channel $b2$.

\subsubsection{Mitigation through Normalization}

Taking the opposite role of bias --- reducing potential alternative information channels --- there are mitigative efforts incorporated in some channels. Just like with potential biases we cannot assume these are effective without analyzing data, but we can note their presence and how their effects change downstream effects, just like with bias. Most importantly we can use the presence of potential bias and mitigation to analyze \emph{structural effects}, which can inform design, assessment and addition of additional mitigative efforts.
There are three channels with explicit mitigative purpose in the modeled information flow. Firstly we have the AI assisted creation of the job description in channel $(a)$. Here the tool includes a bias detection tool working on the language of the job description. Further, with the aim of reducing gender biases both candidates and job descriptions go through a normalization process in channels $(b1)$ and $(b2)$ to remove gendered language from the comparison elements $EC$ and $ER$. From the perspective of IFM, these normalization channels can be seen as targeted information loss --- to the degree they are effective. This means a gendered-language sub-feature(s) $lg \times (DB, CV)$ are lost from downstream processing after the normalization to $EC$ through $(b1)$. This does not mean all potential proxies of gender is removed, but in order for them to remain, it must be possible to show an information path from a feature $g$ to a proxy which is present in $EC$. We might also keep in mind that key downstream channels, such as $(b3)$ operate particularly on language (through semantic distance comparison between features of $EC$ and $ER$).

This concludes the construction of the information flow models for the use-case, and have covered \textbf{Steps 1-4} in the IFM methodology. In the next section we are going to show how this assist further analysis of stakeholder impact.

\section{Analysis and Discussion} \label{sec:discussion}
In the previous section an information flow model spanning the use case and it's context was presented. Here we will explore how this structure can be used to analyze not just the properties of the decision making itself, but also downstream effects on stakeholders, following \textbf{Step 5} of the IFM methodology. 

\subsection{Analysis of Impacts}

While Biases are properties of the decision process, Impacts must be considered in relation to the affected stakeholders or subgroups thereof. In this analysis, we will primarily consider the candidates and to what degree they might be unfairly treated or discriminated on protected characteristics or otherwise.
In order to do so, we begin with a list of possible discriminatory outcomes. and consider if there are paths through the information flow model which contribute to these outcomes.

\begin{itemize}
    \item \textbf{O1: Gender-discriminatory hiring outcomes} --- An outcome in which one gender is systematically underrepresented in the final candidate list $C4$, despite equal underlying qualifications.
    \item \textbf{O2: Exclusion of candidates with non-standard CVs or career paths} --- Candidates with informal, nonlinear, or atypical backgrounds might be disfavored due to mismatches with formal criteria.
    \item \textbf{O3: Penalty for non-native language users} --- Candidates whose CVs or online profiles differ linguistically from standard corpora are mismatched or undervalued during semantic comparison or keyword extraction. Less recognized educational institutions or past employers might not be recognized or valued by scoring mechanisms.
    \item \textbf{O4: Location-based favoritism} --- Favoring of candidates from specific regions possibly linked to protected status, i.e `redlining' through biased scoring heuristics.
\end{itemize}

We now analyze whether the information flow model allows for such outcomes to emerge, and if so, through which pathways. For each outcome, we identify a potential path from specific biases in the system to discriminatory effects at $C3$ or $C4$. These traces represent feasible chains of information distortion or exclusion that are not corrected before reaching decision points that affect candidates.

We will in this analysis take a deeper look at gender-based discriminatory impacts as an illustrative example and then present the result of a similar approach to $O2-O4$. 

\subsubsection*{O1: Gender-discriminatory hiring outcomes}

Gender-based discriminatory effects often originate in pre-existing societal or historical biases getting reinforced by embedding into datasets, models or practices~\cite{weinberg2022rethinking}. In this case, gender-based discriminatory effects may enter through several parts of the process: the \emph{Client}, the \emph{Recruiter}, the \emph{Semantic Model}, or the \emph{Rule-Based Scoring}. For discrimination to occur, gender (or a proxy for it) must be available as information in the flow. In this case, such information can originate in the \emph{CV} and in personal interactions with the candidate ($C1$, $C2$, $C3$). 

Using the IFM, we can see that there are several possible pathways from these potential sources towards the end result. Each of these paths form a potential scenario for impact (where the bias changes the outcome). Below we consider them in turn:

\begin{itemize}

    \item \textbf{Client bias.} If the Client holds gender preferences, these may influence requirements $R0$ or the Recruiter’s understanding $R2$. While $R0$ is subject to mitigation through a bias-detector at $(a)$, the Client’s final decision $(h)$ remains outside of control. Thus, even with a neutral process $(a$–$g)$, discrimination may still arise in the last step.  
     
    \item \textbf{Recruiter bias.} If the recruiter is biased in their assessment of vacancy requirements, this will manifest as a biased $R2$. This might for instance be caused by a recruiter trying to meet assumed biased requirements of the Client, real or not. A biased $R2$ affects all downstream steps $(e)$, $(f)$, $(g)$, since each evaluation of the candidate is conditioned on $R2$.  
	
	

    \item \textbf{Semantic model bias.} Societal or historical gender associations may be embedded or enhanced by semantic distance models. Here, however, in extraction for use in the AI Matching both candidate information and vacancy-requirements ($CV \rightarrow EC$, $R0 \rightarrow ER$) undergo a normalization of gendered terms and language. Conditional on the effectiveness of this mitigation, the path from semantic bias to outcomes is closed.  
    
   \item \textbf{Rule-based scoring bias.} Even after a language based normalization, proxies such as gaps in employment history could encode gender differences (e.g., due to parental leave). In this case however, interviews revealed that the AI matching does not evaluate or capture dates or durations of employment history. This means information on employment gaps is not present. We therefore treat this path as closed.

\end{itemize}

\textbf{Conclusion:} \emph{Conditional on the mitigation}, since there are no unmitigated paths, we see no reason to believe the AI Matching is discriminatory on gender. 
There are, however, several direct routes to gender-based discrimination through the Recruiter and, externally, the Client --- and these paths are entirely unaffected by the three mitigative efforts.

\subsubsection*{O2: Exclusion of candidates with non-standard CVs or career paths}

Nonlinear or atypical career paths could in principle disadvantage candidates if the matching system penalizes gaps or deviations from standard trajectories. However, in this case the AI matching does not evaluate time or continuity of employment. Instead, non-standard paths simply provide more opportunities for positive matches across skills and experiences. Thus, no structural path from career irregularities to discriminatory impact is identified within the AI matching. As before, possible bias remains through the Recruiter’s interpretation ($R2$) or the Client’s evaluation $(h)$.

\subsubsection*{O3: Penalty for non-native language users}

Candidates using non-standard linguistic forms or with CVs from less recognized institutions might be disadvantaged if the semantic model fails to normalize input or if employers or educational institutions are weighted against each other. Here, however, the extraction $CV \rightarrow EC$ do not capture employer or institution names, merely job-titles. No active path of discrimination can therefore be found within the AI matching itself. Bias could still enter through the Recruiter’s interpretation ($R2$) or at the Client’s final decision stage.

\subsubsection*{O4: Location-based favoritism}

Location information plays a special role in the rule-based scoring, where distance to the employer directly modifies the weighted score. This favors candidates from metropolitan or industrial regions over those from less populated areas. While not discriminatory on protected features in itself, the structural effect of distance based scoring can intersect with socio-economic or demographic patterns, leading to geographically correlated disadvantage e.g redlining-like effects. Unlike $O2$ and $O3$, this path remains open within the AI matching and may contribute to location-based discriminatory impacts. Of special note is that this location based scoring is by design, even though potential structural interaction with demographic patterns are not.




\subsection{Summary of Impact Paths}

Three fundamental impact pathways are detected within the IFM, which are not affected by mitigating efforts. Overlapping paths relating to several outcomes have been combined.

\begin{align*}
\textbf{I1:} & \quad C3:\textit{Presentation} 
   \xrightarrow{\textbf{Biased Client (h)}} 
   C4 && O1, O2, O3 
   \\[6pt]
\textbf{I2:} & \quad \textit{Vacancy} 
   \xrightarrow{\textbf{Biased Recruiter (a)}} 
   [\textit{Biased R2}] \quad 
   C1, [R2] \xrightarrow{(g)} C2,[R2] \xrightarrow{(h)} C3 
   \xrightarrow{\text{Client}} C4 && O1, O2, O3 
   \\[6pt]
\textbf{I3:} & \quad \textit{EC, Weighted score} 
   \xrightarrow{\textbf{Location Multiplier}} 
   C0 \xrightarrow{(e)-(h)} C4 && O4
\end{align*}

These downstream impacts \textbf{I1-I3} are noted in Table \ref{tab:transitions} and \ref{tab:ai_match_breakdown} for each channel.

The above argument around outcomes $I1-I3$ and $O1-O4$ does \emph{not} mean that there are no other harmful or discriminatory outcomes.







\subsection{Reflections on the Analysis}

The analysis highlights, in line with~\cite{selbst2019fairness,weinberg2022rethinking} that fairness in deployed systems is not solely a matter of local model-properties but an emergent property of how information is transformed, abstracted, and acted upon across organizational layers and ultimately how it actually affects stakeholders. The case illustrates that mitigation embedded within the AI system may successfully remove some paths of discrimination, while parallel pathways might reintroduce or maintain harmful patterns.

With regards to the AI system the analysis revealed only one - unmitigated - direct path from the AI Matching to discriminatory outcome: the location based scoring modifier. This heuristic strongly favors candidates living close to employers, which can correlate with socio-economic or demographic segregation. Our recommendation from the study was therefore to re-examine this rule by either removing it, tailoring it to vacancy type and location, or testing its effects through simulations on synthetic or real data. This would help avoid unjustified forms of localization bias.

However, the analysis points to risks of an explainability mismatch. If the AI tool only provides a list of candidates, recruiters may treat it as authoritative without questioning its heuristics. Interviews with recruiters show several misunderstandings on how the tool actually operates. A more transparent strategy would be to provide recruiters with the tool’s sub-scores alongside the answers. This would allow them to, if desired, critically evaluate AI results, spot errors, and align their understanding of the automated outputs with their own professional judgment. It would also reduce opacity in accountability pathways by making explicit how particular features contribute to candidate evaluation.

While the analysis found most pathways of discrimination in the AI Matching blocked by mitigation, it was also showed that these mitigative mechanisms within the AI system, such as normalization steps, remain largely ineffective to address potential social sources of bias. This includes the bias check on $R0$ which was explicitly designed to capture social biases, but is largely sidestepped by recruiters’ intuitive understanding of requirements ($R2$).

$R2$ can still carry and reproduce biased assumptions, regardless of the AI Matching or mitigation. To counteract this, socio-technical mitigative measures are necessary. Guidelines on gender equality in the selection process, supported by follow-up monitoring or logging, could strengthen accountability and embed mitigation within organizational practice. Structuring recruiter reasoning and tracking decision patterns over time may further support fairness monitoring and, ultimately, more equitable outcomes.

While this was outside the scope of this paper and not explored in the above analysis, the IFM model can also be used to show that not all decision steps carry equal weight for the final outcome. From the information available at each step we can strongly suspect that the channel \emph{(e) Recruiter selection} and the channel \emph{(h) Client selection} will carry stronger weight than the other steps, just because these channels are less structured with a larger variation and with many alternatives - detected and not detected. These steps will therefore combine high decision weight and low accountability. It might be hard to address the behavior of the client, but it is possible to put additional structure or follow-up mechanisms around channel \emph{(e)}.

\subsection{Validation and Limitations}

The case study also provided validation of the IFM framework. When presented back to use case-holders, the model was perceived as accurate and illuminating, clarifying their own processes better than existing documentation. Stakeholders reported that they intend to use the material for internal education and for highlighting socio-technical origins of bias. In post-study reflections, the model was also seen as a useful pedagogic tool, capable of raising awareness and structuring internal discussions about fairness and accountability. In our validation interviews, it has also surfaced that stakeholders felt the participatory method itself was both helpful and educational in that it created reflection and overview, even before seeing the final model. These reactions confirm that IFM can support not only technical redesign but also organizational learning and governance.

At the same time, the limitations of the study must be acknowledged. Most notably, candidates were not interviewed, meaning that the analysis of impacts is necessarily more speculative than the modeling of decision structure, bias, and mitigation. The IFM model also remains interpretive, depending on the fidelity of stakeholder accounts and the perspective of the modeler. Feedback from external stakeholders highlighted the importance of explicitly attending to intersectional discrimination and modeler bias, suggesting that additional methodological scaffolding is needed to ensure robustness and inclusivity in future applications.

\subsection{Situatedness and Participation}

A recurring critique of AI ethics is the tendency to produce abstract taxonomies of fairness problems without sufficient connection to practice~\cite{rain,green_algorithmic_2020}. Fairness metrics, bias taxonomies, and governance templates argue for high-level principles in the abstract but fail to specify how such issues manifest in concrete socio-technical contexts~\cite{selbst2019fairness,weinberg2022rethinking,metcalf2021algorithmic}. One of IFM’s central strengths is that it situates fairness concerns within the concrete structure of decision-making. By modeling how information flows between technical components and human actors, the framework makes clear that not all general problems are equally contextually relevant: some are structurally blocked, others tied to specific decision points and actors. This contextualization both narrows the problem space and connects, or dissolves, abstract concerns to concrete impacts and mitigation strategies.

The participatory process of building the IFM model is key to this situatedness. Each stakeholder brings a partial and localized perspective — developers know technical functions, recruiters know their practices, external representatives know broader risks — and these fragments are connected through the modeling process into a coherent structural account. Thus, the resulting model embodies both the situated perspectives of stakeholders and the structural relationships between decision steps. This dual role allows the model to function as a shared artifact for reflection, bridging the gap between abstract fairness discourse and concrete socio-technical practice.
Taken together, the case study shows how IFM bridges the abstraction gap through its \emph{structural} representation of decision-making, its \emph{holistic} coverage of technical and social components, and its grounding in \emph{situated} stakeholder perspectives.

\section{Conclusion} \label{sec: conclusion}
This case study demonstrates how information flow modeling can be used to identify, trace, and contextualize fairness risks in socio-technical decision processes. Applied to a real-world recruitment setting, the framework made visible how biases may emerge and propagate across both automated and human decision steps. By explicitly modeling information structure and transformation, it enabled a unified analysis of technical mechanisms and organizational practices—offering a practical tool for increased socio-technical transparency and contextually grounded fairness assessment.

Beyond the case itself, IFM complements traditional algorithmic performance and fairness tools by providing the contextual layer often missing in real-world applications. Its structural, holistic, and situated approach helps connect abstract fairness concerns to concrete system behaviors and stakeholder impacts.

Looking ahead, IFM lends itself to further development in several directions. Scenario modeling with alternatives could support exploration of what-if configurations and their consequences. In addition, formal representation of bias and mitigation as structural substitutions would enable systematic comparison of system designs and mitigation strategies. Together, these extensions would strengthen IFM’s role in comparative audits and compliance evaluations, and link fairness analysis more directly to risk- and safety-oriented assessments under frameworks such as the EU AI Act.

\begin{acknowledgments}
    This work was supported by the “AEQUITAS” project funded by European Union’s Horizon Europe research and innovation programme under grant number 101070363. The views and opinions expressed are those of the author(s) only and do not necessarily reflect those of the European Union or the granting authority.
\end{acknowledgments}

\section*{Declaration on Generative AI}

 During the preparation of this work, the authors used GPT-5 in order to: Paraphrase and Reword \& Improve Writing Style. After using these tools, the authors reviewed and edited the content as needed and takes full responsibility for the publication’s content. 

\newpage
\bibliography{ref}



\end{document}